\documentclass[preprint,12pt,authoryear]{elsarticle}

\usepackage{amssymb}
\usepackage{booktabs}
\usepackage{amsmath}
\usepackage{multirow}
\usepackage{url}
\usepackage[hidelinks]{hyperref}
\usepackage{lscape}
\usepackage{subcaption}
\usepackage{tabularx}
\usepackage{psfrag}% http;//ctan.org/pkg/psfrag
\usepackage{balance}
\usepackage{xcolor}
\usepackage{enumitem}
\usepackage{nicefrac}
\usepackage{xurl}

\newcommand{\revision}[1]{\textcolor{black}{#1}}

\newcommand{\pquote}[2]{\textit{`#2'}~(#1)}

\newlength{\tablecaptionspacing}
\journal{International Journal of Human-Computer Studies}

\begin{document}

\begin{frontmatter}



\title{The Experience of Running: Recommending Routes Using Sensory Mapping in Urban Environments}


\author[label1]{Katrin H\"ansel}
\author[label2]{Luca Maria Aiello}
\author[label3, label6]{Daniele Quercia}
\author[label4]{Rossano Schifanella}
\author[label5]{Krisztian Zsolt Varga}
\author[label6]{Linus W. Dietz}
\author[label3]{Marios Constantinides}

\affiliation[label1]{organization={Yale University},
            city={New Haven},
            state={Connecticut},
            country={United States}}

\affiliation[label2]{organization={IT University of Copenhagen},
            city={Copenhagen},
            country={Denmark}}

\affiliation[label3]{organization={Nokia Bell Labs},
            city={Cambridge},
            country={United Kingdom}}

\affiliation[label4]{organization={University of Turin},
            city={Turin},
            country={Italy}}

\affiliation[label5]{organization={Nokia Bell Labs},
            city={Budapest},
            country={Hungary}}

\affiliation[label6]{organization={King's College London},
            city={London},
            country={United Kingdom}}
            
% \affiliation{organization={Nokia Bell Labs},%Department and Organization

\begin{abstract}
Depending on the route, runners may experience frustration, freedom, or fulfilment. However, finding routes that are conducive to the psychological experience of running remains an unresolved task in the literature. In a mixed-method study, we interviewed 7 runners to identify themes contributing to running experience, and quantitatively examined these themes in an online survey with 387 runners. Using Principal Component Analysis on the survey responses, we developed a short experience sampling questionnaire that captures the three most important dimensions of running experience: \emph{performance \& achievement}, \emph{environment}, and \emph{mind \& social connectedness}. Using path preferences obtained from the online survey, we clustered them into two types of routes: \emph{scenic} (associated with nature and greenery) and \emph{urban} (characterized by the presence of people); and developed a routing engine for path recommendations. We discuss challenges faced in developing the routing engine, and provide guidelines to integrate it into mobile and wearable running apps.
\end{abstract}


% \begin{highlights}
% \end{highlights}


\end{frontmatter}


% \section{}
% \label{}
\section{Introduction}
In a run, sensory encounters (e.g., visual cues, smells,  and sounds)  impact the feeling of pleasure, safety, and freedom of runners. Green spaces are associated with low stress~\citep{Tyrvainen2014-stressurban}, good cardiovascular functioning~\citep{lanki2017-urbanheart}, and mediated vitality~\citep{ryan2010-vitaliygreen}.
Suggesting more pleasant and beautiful running routes might well improve the running experience. 
Finding good running routes is not easy. Tools such as Strava~\citep{strava}, Runkeeper~\citep{runkeeper}, or MapMyRun~\citep{mapmyrun} are frequently used to build running routes. However, these tools strictly focus on routes' qualities such as distance, elevation gain, or surface, rather than the properties of routes' surroundings.
There are clear shortcomings with these tools in terms of adaptability and personalization. For example, their routing features are tailored to training instead of recreational running. Instead, automatically generated routes should be based on more contextual information. Supported by the increasing availability of open data on the urban environment,
computational running routes may provide an improved, safer, more pleasant experience catered to personal preferences.  Furthermore, social media may be successfully used for geographic profiling, not least because it has already been used to geo-spatially identify environmental and sensory properties of a place such as beauty, smell, or sounds~\citep{quercia2014-happymaps, quercia2015-smellymaps, aiello2016-chattymaps}, and to even capture emotional perceptions~\citep{Resh2016-urbanemotion}.

The specific problem with running is that its psychological experience is still not fully understood, and finding good running routes, specifically tailored to support a positive experience is not easy. Common strategies focus on using community-generated routes or personalized ones derived from past experience. Especially the latter approach is often time-consuming and involves trial-and-error. 
Community-generated routes, on the other hand, are not customizable, may not be available in the targeted area, and are not tailored to personal preferences or temporal circumstances. There is little work on automatically generating pleasant, safe, and personalized running routes based on environmental and sensory features. Providing more pleasant and personalized routes not only can support optimal running performance but also can lead to more fulfilling experiences. To this end, we formulated three research questions (RQs):

\begin{enumerate}
    \item[\textbf{RQ\textsubscript{1}}:] What characterizes the experience of runners?
    \item[\textbf{RQ\textsubscript{2}}:] What factors---especially environmental factors---influence the experience and how can these be mapped to quantifiable sensory and environmental properties?
    \item[\textbf{RQ\textsubscript{3}}:] How could we effectively measure this experience? 
\end{enumerate}

Our methodology adopts an exploratory sequential design~\citep{fetters2013achieving}, which is a mixed-method approach that combines qualitative methods for initial exploration followed by quantitative methods for generalization and quantification of findings~\citep{Creswell:2006-mixed-methods, fetters2013achieving}. Initially, we conducted an interview study with 7 participants to gain insights into runners' perceptions, experiences, route selection, and use of running technologies. From these interviews, we identified five themes characterizing the running experience, which guided the design of an online survey. The online survey aimed to generalize the findings from the interviews (RQ\textsubscript{1}), quantify environmental and sensory aspects supporting a positive experience (RQ\textsubscript{2}), identify main components of the running experience for a short experience sampling questionnaire (RQ\textsubscript{3}), and generate two types of routes: \emph{scenic} (associated with nature and greenery) and \emph{urban} (characterized by the presence of people). Based on these two types of routes, we developed a routing engine for scenic and urban path recommendations. In so doing, we made three main contributions:

\begin{enumerate}
    \item We identified 5 key themes through semi-structured interviews with 7 regular outdoor runners (\S\ref{sec:interviews}), which then informed a large-scale online survey.
    \item We administered the large-scale survey with 387 participants (\S\ref{sec:survey}), performed a Principle Component Analysis of its responses and determined the 3 questions that best capture one's pre- and post-run experience in a short and compact manner---the outcome is an \emph{Experience of Running Scale (ERS)}. We extracted two path preferences with a specific environmental and sensory profile---\emph{scenic} and \emph{urban}---that emerged from the survey responses, and found associations with personality traits. 
    \item We implemented a routing engine incorporating scenic and urban path preferences (\S\ref{sec:design}). The routing is driven by scores of seven dimensions,
 which are weighted according to their importance discovered in the survey.
\end{enumerate}

We conclude by discussing open challenges in designing and evaluating the routing engine, and provide guidelines to integrate it into mobile and wearable running apps (\S\ref{sec:discussion}).
\section{Related Work}
Running is an enjoyable and intrinsically appealing physical activity with a low entry barrier and without the need for special facilities or equipment. Individuals who engage in frequent running enjoy a range of positive effects such as health benefits related to a general increased physical activity \citep{Running-health}, elevated mood \citep{mood-run}, and even an antidepressant effect \citep{BRENE2007136,MalchrowiczMosko2018Running}. Sports psychology identified a number of motivations for running, including physical health (weight loss~\citep{LeonGuereno2021Motivation,Running-health}), mental benefits (e.g., well-being~\citep{Popov2019Relations},  self-esteem~\citep{LeonGuereno2021Motivation,Running-health}, discomfort experienced when a run is missed~\citep{Carmack_CRscale}), and  social aspects (e.g., affiliation~\citep{LeonGuereno2021Motivation},  and the goal of achieving the body image of a `running body'~\citep{Running-health}.) External factors are also important. One may engage with running to gain recognition for athletic achievements~\citep{LeonGuereno2021Motivation} or being active in nature~\citep{Deelen2019Attractive}.
Overall, the focus of prior research was mainly on health benefits~\citep{Running-health}, running performance, and social aspects \citep{social-run}. Yet, the psychological experience of running remains an unresolved task in literature.
 
In what follows, we surveyed various lines of research that our work draws upon, and grouped them into three areas: \emph{i)} the sensory experiences of running in a city, \emph{ii)} commonly used mobile computing technologies that support health and exercise, and \emph{iii)} technologies for running route recommendations.

\subsection{Sensory Mapping of Cities}
The running experience may vary based on external factors such as terrain, weather, and sensory stimuli. Exercise in natural, `green' spaces has been found to reduce perceived stress and promote overall well-being, shifting focus away from internal negative feelings~\citep{bowler2010-naturalexercise, harte1995-runnermood}. For example, many runners tend to opt for beautiful and scenic routes to distract themselves from the challenges of running~\citep{Morgan-associ,Deelen2019Attractive}.

Previous research on running has focused primarily on understanding affective experiences during runs through experience sampling methods~\citep{Jogchalking}. \citet{urban-cycle-stress} explored stress factors among cyclists, constructing stress maps from sensor data and revealing correlations between noise, cycle paths, and time of day with stress responses~\citep{urban-cycle-stress}.
Nonetheless, there is limited research that fully captures the running experience.

When it comes to capturing sensory perceptions of people's environment, previous research mostly utilized crowdsourced social media data to estimate factors such as smell, sounds, and visual beauty. For example, street-level city sounds were captured using image metadata from Flickr, with millions of images analyzed using computer vision to categorize sound tags such as transport, mechanical, human, music, nature, or indoor~\citep{aiello2016-chattymaps}. Similarly, social media data from platforms such as Flickr, Instagram, and Twitter was employed to map urban smellscapes~\citep{quercia2015-smellymaps}. These datasets have been used to recommend pleasant paths based on crowdsourced user preferences~\citep{quercia2014-happymaps}, with the datasets accessible for London via \url{http://goodcitylife.org}. In this work, we explore sensory mapping data sources to understand people's perceptions of the urban running environment.

\subsection{Mobile and Wearable Apps for Runners}

Personal informatics for health has gained significant traction~\citep{Motahar2022Review,Epstein2020Mapping}. Mobile and wearable apps offer a variety of features such as tracking physical activity, nutrition, and sleep~\citep{yfantidou2021self, park2020wellbeat,Lee2022Understanding,menheere2020runner}.
For example, the \emph{MyFitnessPal} app allows users to monitor food intake and exercise, while the \emph{Sleep Cycle} tracks sleep patterns~\citep{myfitnesspal,sleepcycle}. Apps like \emph{Strava} and \emph{Nike Run Club} cater specifically to runners, provide features for monitoring distance and pace~\citep{strava,nrc}. 
However, these apps have shortcomings because they tend to focus on health and performance aspects without considering user preferences (e.g., preferred running paths). None of the aforementioned apps use urban theories to improve the running experience.

Another area of research explores apps designed to alter user behavior~\citep{rabbi2015automated} and boost motivation~\citep{babar2018run, menheere2020runner,Mulas2011Everywhere}. These apps often employ techniques such as gamification~\citep{deterding2012gamification} and social support to inspire regular physical activity. For example, \emph{Zombies Run} is a game that uses a narrative about the zombie apocalypse, motivating users to continue running, while \emph{Carrot Rewards} rewards users with points for meeting physical activity targets~\citep{zombiesrun,carrotrewards}. While effective at encouraging exercise, these behavioral change apps, which often use gamification techniques, require ongoing support and engagement to sustain impact. Therefore, gamification techniques may not resonate with all users, highlighting the need for diverse approaches to promote physical activity.

\subsection{Route Recommendation in Cities}

Selecting the right route greatly impacts the running experience, but finding convenient routes that match a user's needs can be challening. Ideally, these routes should minimize obstacles such as traffic lights, offer exposure to nature, and be safe during low-light conditions~\citep{Deelen2019Attractive}.
Commercial apps like \emph{Runkeeper}~\citep{runkeeper}, \emph{MapMyRun}~\citep{mapmyrun}, and \emph{Strava}~\citep{strava} allow users to share and generate personalized routes based on criteria such as terrain and popularity. However, these recommendations often lack personalization, relying on basic features such as hilliness or road surface. In sports route recommendations, \citet{McGookin2015-runnerscrowdsource} used crowd-sourced data to assess areas suitable for running~\citep{McGookin2013-runnerfoursquare, McGookin2015-runnerscrowdsource}, focusing on areas rather than paths. \citet{run-generation} used street properties and points of interest from OpenStreetMap to generate running paths~\citep{run-generation}. \citet{SU2010495} explored enjoyable cycling routes based on factors like air pollution and safety~\citep{SU2010495}, while \citet{Zeile2016-urbanemotion} combined social media emotions with wearable affect sensing to recommend cycling routes~\citep{Zeile2016-urbanemotion}.

Regarding general route recommendations in urban settings, several studies addressed the tourist trip design problem, that is, to optimize traveler experiences within limited time~\citep{Gavalas2014Survey,Vansteenwegen2007Mobile,Herzog2019Tourist}. \citet{Herzog2019Integrating} proposed an approach incorporating route attractiveness attributes such as trees, pollution, and cleanliness derived from OpenStreetMap data~\citep{Herzog2019Integrating}. Additionally, \citet{quercia2014-happymaps} pioneered recommending pleasant walkable paths in cities based on sensory mapping from diverse data sources~\citep{quercia2014-happymaps}. Despite these efforts, there has been little focus on computationally-generated route recommendations tailored for runners based on nuanced environmental and sensory features relevant to their activity. Our research delves into the urban running experience, particularly how it can be characterized and how it can inform the design of a running route recommendation engine based on desired route properties.
\section{Qualitative Exploration of the Running Experience}
\label{sec:interviews}

To explore the experience during runs, we conducted semi-structured interviews with $7$ outdoor runners. Next, we explain our method and procedure (\S\ref{subsec:procedure}), the factors influencing running experience (\S\ref{subsec:factors_experience}), and discuss the implications for the design of a survey that captures running experience quantitatively (\S\ref{subsec:implications_survey}).

\subsection{Method and Procedure}
\label{subsec:procedure}
The interviews followed a semi-structured approach~\citep{bernard2011research}, where the interviewer followed an interview guide with proposed questions and topics to cover~\citep{bernard2011research}. The interviews were audio-recorded and transcribed with annotations from the ATLAS.ti software.\footnote{\url{https://atlasti.com/}}
The annotation was followed by a thematic analysis adopting a process of developing from lower level `codes' in a first annotation round towards higher level `themes'~\citep{braun-thematic}. \revision{Although the qualitative sample consisted of 7 participants, this number was sufficient to achieve thematic saturation. Given the richness and consistency of the data collected, we concluded that additional interviews were unlikely to generate new insights~\citep{braun-thematic}.}

\begin{table}[t!]
	\centering
	\caption{Demographic information and running habits of the study participants}.	\label{tab:interview-demographics}
        \scalebox{0.8}{
	\begin{tabular}{llllll}
		\toprule
		\textbf{ID} & 	\textbf{Age}& \textbf{Gender} &\textbf{Years Running}                                                                                                                                                                                                   & \textbf{Weekly Milage} & \textbf{Running Area} \\ \midrule 
		\textbf{P1} & 37& M               & 15                           & 
		25-30 km       & urban   \\
		\textbf{P2} & 42& M               & 5                                                                                                                                        &  $\sim$ 20 km & parks and roads         \\
		\textbf{P3} & 29& M               & 25                           & 15 km  & gyms and parks               \\ 
		\textbf{P4} & 42& M               & 5                                                                           & 10-20 km  & parks              \\ 
		\textbf{P5} & 25& F               & 5                                                                             & 15-20 mi & parks          \\ 
		\textbf{P6} & 26 & F               & 8                            & 40 km    & parks, forests             \\ 
		\textbf{P7} & 29& F               & 11                  
		& 12 km   & urban areas, sometimes parks                \\
		\bottomrule
	\end{tabular}
 }
\end{table}

\subsubsection*{Participant Profiles}

Participants were recruited through word of mouth and mailing lists, comprising a convenience sample of regular outdoor runners who ran at least twice a week in the past three months. They were informed that the study aimed to explore the experiences of urban runners and provided informed consent, including the acknowledgment of audio recording for transcription. Interview recordings totaled 4:15 hours, and demographic and running behavior data are summarized in Table~\ref{tab:interview-demographics}. Interviews were conducted face-to-face (2) and online via a video conferencing tool (5).

\subsubsection*{Interview structure}

The interview guide included questions on motivation, preparation, path choices, typical experiences during and after a run,  and criteria for an ideal running route. The final section focused on current technology usage and reflections to guide potential improvements regarding routing for runners. An outline of the interview topics and questions is provided in the Table~\ref{tab:interview-guide} in the Appendix.

\subsection{Factors Influencing the Running Experience}
\label{subsec:factors_experience} From the thematic analysis, five themes emerged: \emph{environmental experience}, \emph{focus}, \emph{abilities}, \emph{post-running experience}, and \emph{path choice strategies}.
\smallskip

\noindent\textbf{Environmental Experiences. }Six participants, excluding P3, expressed their appreciation for natural elements such as greenery during runs, which are known to contribute to well-being and positive health outcomes~\citep{bowler2010-naturalexercise}. They noted that natural spaces offer cleaner environments, better air quality~(P2, P4), and distance from traffic~(P4). Conversely, the urban environment was described as \emph{`fun'}~(P1) and as a means to experience the city (P1), with P6 emphasizing the beauty of running in urban settings. However, selecting the appropriate urban environment for running was highlighted as important~(P2). Safety concerns and urban design elements such as running surfaces and obstacles were also discussed. P3 emphasized the importance of feeling safe in their running location, echoed by P5 and P7 \pquote{P5}{So I want to know there will be people. Partly for safety, because it is nice to know there is [sic] people.}. Smooth running surfaces were preferred by P2 and P7 for performance and injury prevention. Environmental obstacles like traffic lights and pedestrians were mentioned as sources of frustration by five participants, with perceptions of co-habitation being seen as either frustrating or enhancing the perceived safety.
\smallskip

\noindent\textbf{Focus during the Run.} Although not directly prompted in the interview guide, all participants discussed key elements they focus on during their runs, including music, bodily sensations, internal dialogue, and nature. Music emerged as a significant motivator for all participants, serving either as a source of motivation (P5, P7) or as a means to \pquote{P2}{pass the time}. However, P1 noted that music could interfere with thinking to some extent. Thus, music was perceived as a distraction from the challenges of running, enhancing enjoyment for some (P7), but not preferred by all participants. For example, some opted for silence to engage with their own thoughts (P5) or to immerse themselves in their surroundings (P4). An external dissociative strategy mentioned by P3 involved using the Nike running app with audio training, which helped distract him from fatigue, enabling longer runs. Conversely, participants highlighted internal focuses during their runs, including their thoughts (P1, P4, P5), bodily sensations (P1, P2, P3, P4, P6), breathing (P1, P3, P6), and pain (P4). P2 emphasized paying attention to bodily sensations to \emph{`gauge my fitness from what I feel is lacking.'} Running was described as a meditative activity by P4, providing \emph{`space and time to process some things which are happening in your life.'} Engaging with the surroundings emerged as another positive distraction during running. Participants appreciated scenic routes by canals (P7) and natural elements like autumn colors (P4), which diverted attention away from running.
\smallskip

\noindent\textbf{Abilities and Struggle.} Running was metaphorically linked to a roller coaster ride by P6, who described its fluctuating nature, a sentiment echoed by other participants (P5, P6). P7 highlighted the changing bodily sensations during a run, noting a shift in breathing patterns and a feeling of lightness after the initial difficulty. Struggles and hardships included physical challenges like pain, fatigue, and breathing difficulties, as well as unpredictable bodily sensations (P4). These obstacles could impede runners from achieving their distance or pace goals. P2 mentioned struggling with new routes or running familiar routes in reverse, while P6 acknowledged the necessity of challenging workouts for improvement. Fear of not finishing a run was cited as demotivating and detrimental to confidence (P3). Participants employed various coping strategies to counter fatigue and adversity. P3 found diverse and interesting environments, while P5 and P7 found music helpful. P6's coping strategy was to rely on social support: \emph{`When it gets painful and tired and I want to give up, I usually think of my friends and family, people I love, and this helps me to push through it.'} 
\smallskip

\noindent\textbf{Post-Run Experience.} The way our participants felt after their runs was actually their main motivation to run in the first place. The feelings of \emph{freedom} and \emph{clear mind} were reported by several participants (P1, P4, P6, P7). Further, one of them reported feeling more chatty and \emph{socially connected} (P1). The feeling of \emph{achievement} (of being proud) also emerged. P4 shared it with his partner: \emph{`We talk about it with my partner as well. I say: "today I am very proud of myself."'} Finally, being hungry after a run was reported by P4, P6, and P7. 
\smallskip

\noindent\textbf{Path Choice Strategies and the Optimal Running Routes.} Some participants (P1, P7) expressed a preference for spontaneous path choices, finding enjoyment and pleasure in exploring new places or cities, particularly while on holiday (P1, P2). During such runs, metrics like distance, speed, and pace were of secondary importance (P5), with the focus shifting towards the experience rather than performance. P2 described exploratory fun runs as non-goal-oriented, focusing instead on observing performance and bodily reactions to different situations. Tailoring routes to specific goals was a common practice among participants. Some routes were designated as training routes aimed at achieving targets, while others were recreational. P4 highlighted: \emph{`It also depends if I am in the mood to push myself to run a difficult route or if I just need to relax.'} Participants often sought recommendations for running routes from other runners, utilizing social features in apps like Strava (P4, P5, P6), although negative experiences, such as overcrowded routes  \pquote{P5}{full of tourists}, were also reported in this context. Recommendations from friends (P6) or participation in running clubs (P5) were also mentioned. Environmental factors played a crucial role in selecting running routes, with considerations including running surface, air quality, and population density (P2). Adjusting routes to training targets involved mental calculations of route segments (P2, P4, P7) or running routes in reverse (P2).
	
All participants utilized apps to log their runs for tracking basic statistics like pace and distance. These statistics were predominantly used to assess fitness levels and monitor progress. P4 also utilized them to analyze potential causes for injuries such as over-training, with P6 using specifically the Endomondo tags to monitor terrain conditions and track the usage of her running shoes. While most participants were diligent in logging and reflecting on their runs (P2, P5, P6, P7), others reported abandoning them altogether and opting to run without their phones (P3 and P4).

\subsection{Implications for the Design of a Quantitative Survey on the Running Experience}
\label{subsec:implications_survey}

As the interviews showed, there is no uniform experience. While some runners preferred a spontaneous experience and actively used running as a tool for exploration~(P1), others were hesitant in having set out routes (P2). 
Further, motivations for running differed: P2 and P3 did it for its health benefits and to reach desired performance levels, while P1 and P4 simply run for enjoyment. Key themes associated with the surroundings included the contrasting nature of natural \emph{vs.} urban spaces, air, and ground quality levels, and obstacles causing a runner to slow down or even stop. 

Motivation, running goals, and personal preferences clearly shape how one runs, how the run itself is perceived, and what is considered a good experience. Within the follow-up survey, we further investigated these aspects by having questions concerning five main themes: \emph{environmental experience}, \emph{focus}, \emph{abilities}, \emph{post-running experience}, and \emph{running path preference}.
We visualized the emerging themes and sub-themes in Figure~\ref{fig:experience-themes} in the Appendix.

\section{Online Survey on Experiences and Path Preferences of Outdoor Runners}
\label{sec:survey}

The design of the online survey was grounded in the resulting themes of the previous interview study. 
The main aim was to explore the running experience in a more generalizable manner, develop a scale for measuring experience, and extract quantifiable environmental and sensory attributes for a good running route.

\subsection{Questionnaire Design}
The first part of the survey focused on demographic information and running behavior. This part included questions on age, gender, ethnicity, five personality traits as per the 10-item BFI-10 scale~\citep{Rammstedt_BFI10}, and employment. It also had questions about the typical run, preferred time of the day for a run, the frequency of running, weekly mileage, and other questions coming from the widely-used Commitment to Running Scale~\citep{Carmack_CRscale}. 
	
The second part of the survey focused on a specific run --- the last one. Apart from the time of the last run and the general enjoyment, we assessed the experience based on the themes and sub-themes from the interviews (\autoref{fig:experience-themes}). Each sub-theme was represented by one or multiple questions. The questions were worded as statements to be rated in terms of the level of agreement on a 5-point scale, like, e.g., \emph{`I listened to music.'} or \emph{`I was happy with my pace.'} The decision to inquire about the last run versus a general run made it possible to account for recency bias \citep{thomas-recency}. 
	
The third part focused on what makes a run ideal in terms of environmental and sensory experiences. The questions were based on five environmental themes: \emph{running ground}, \emph{safety}, \emph{habitation}, \emph{traffic}, and \emph{obstacles}. We further added questions on the sensory perceptions of \emph{smell}, \emph{sound}, and \emph{visual beauty}, not least because environmental elements emerged as very important factors during the interviews. 

\subsection{Participants and Recruitment}
The survey was distributed through Facebook among running communities and mailing lists and remained open for two weeks, receiving a total of 387 responses. The initial page of the survey outlined participant requirements (e.g., being a regular outdoor runner). The demographic profile of the participants is presented in Table~\ref{tab:survey-demographics} in the Appendix, aligned with data from large-scale running surveys such as the National Runner Survey~\citep{running-survey}. \revision{The sample primarily consisted of participants aged between 45-54 years (32.8\%) and 35-44 years (20.4\%), with the majority identifying as female (52.5\%) and male (46.5\%). A lower percentage of the participants were 18-24 years old (7.2\%) or 65 years and older (6.3\%). Furthermore, non-binary and other gender identities were minimally represented (0.2\% each). Although this sample reflects a reasonably balanced gender distribution, it is important to acknowledge the potential demographic homogeneity, particularly with respect to age, with a concentration in the middle-age range. Certain groups, such as younger or older runners, could be underrepresented, which could influence the generalizability of our findings.}

\subsection{Validity of the Questionnaire}
The face validity of the survey questions was evaluated through discussions among co-authors, and feedback was obtained from two test participants in a pilot study whose responses were discarded from the analysis. The items rated on a 5-point Likert-scale were designed to be concise, clear, and focused on a single aspect each, such as \emph{``The route was clean.''} Construct validity was assessed using a correlation matrix among the items, revealing meaningful convergence on the underlying constructs. Additionally, Principal Component Analysis (PCA) yielded consistent and meaningful components.

\subsection{Identifying Components of the Running Experience}
\label{subsec:components-running-experience}

\begin{table*}[t!]
\footnotesize
% \small
\caption{Rotated Component Matrix for the 3 derived components of the experience during a run: \emph{performance \& achievement}, \emph{environment}, and \emph{mind \& social connectedness}.}
\scalebox{0.78}{
\begin{tabular}{@{}lccc@{}}
\toprule
  & \begin{tabular}[c]{@{}c@{}}performance \\ \& achievement\end{tabular} & \begin{tabular}[c]{@{}c@{}}perceived\\ environment\end{tabular} & \begin{tabular}[c]{@{}c@{}}mind \& social\\ connectedness\end{tabular} \\ \midrule
  
It was easy to reach my running goal.                                                                        & .776                                                           &             &                                                                  \\
I was happy with my pace.                                                                                    & .728                                                           &             &                                                                  \\
I felt free during the run.                                                                                  & .668                                                           &             &                                                                  \\
I was hardly frustrated during this run.                                                                     & .664                                                           &             &                                                                  \\
I felt confident beforehand that I can reach my goal in this run. & .637                                                           &             &                                                                  \\                                                            
All things considered, how satisfied were you with the run?       & .619                                                           &             &                                                                  \\
The running experience was as I expected it to be.                                                           & .598                                                           &             &                                                                  \\ 
I was happy with my distance.                                                                                & .584                                                           &             &                                                                                                                                    \vspace{.5em}\\

The route was clean.                                                                                         &                                                                & .757        &                                                                  \\
The ground was good to run on.                                                                               &                                                                & .717        &                                                                  \\
The air quality was good.                                                                                    &                                                               & .685        &                                                                  \\
I was pleased with the beauty of my surroundings.                 &                                                                & .639        &                                                               \vspace{.5em}   \\

I felt connected to people.                                                                                  &                                                                &             & .784                                                             \\
My mind felt clear after the run.                                                                            &                                                                &             & .709                                                             \\

\midrule
{\footnotesize component loadings below $.5$ are suppressed. } \\
\bottomrule
\end{tabular}
}

\label{tab:rotated-component-loadings}
\end{table*}

	To identify the main orthogonal dimensions that explain the survey's responses, we performed a PCA on the experience ratings of the last run. PCA \emph{`determine[s] the linear combinations of the measured variables that retain as much information from the original measured variables as possible'}~\citep[p. 275]{Fabrigar_PCA_def}.
	From the 387 survey respondents, we excluded 37 as they reported that their last run was more than 7 days ago (for a possible bias in their recall); this left us with 350 responses and a ratio of 11.6 cases per variable, which exceeds the widely recommended ratio of 1:10 for PCA analysis~\citep{nunally-sample-pca}.
 
	To ascertain that our PCA results are valid, we performed three tests. 
	First, the Pearson correlation matrix was observed. Each item's 5-point agreement rating (ranging from `strongly disagree' to `strongly agree') was previously transformed into a numeric score (ranging from $-2$ to $2$), and subsequently, the correlation analysis was performed on these. The correlation matrix showed significant correlation coefficients above $0.3$ with at least one other item for 24 of the 30 items, suggesting reasonable factorability. 
	
	Second, Bartlett's Test of Sphericity was performed. Bartlett's Test of Sphericity tests the null-hypothesis that the correlation matrix is an identity matrix, which would indicate that the variables of the questionnaire are unrelated and unsuitable for continuation with the PCA. The results of the test were found to be significant ($\chi^2 (350) = 2626.30, p = .000$), supporting the application of the PCA. 
		
	Third, the Kaiser-Meyer-Olkin (KMO) \citep{Kaiser:1947-KMO} measure of sampling adequacy was $.785$, which was above the recommended threshold of $.7$~\citep{lloret-msa-threshold}. A high KMO indicates that the variance is caused by underlying factors.
	Eleven items with a KMO below $.7$ in the anti-image correlation matrix were excluded from the analysis~\citep{normanstreiner2008}. This exclusion resulted in improved values of $\chi^2 (350) = 1610.39, p = .000$ in Bartlett's Test of Sphericity, and in an overall KMO score of $.858$. This made the Principal Component Analysis more robust with the remaining 19 items.

	Within the PCA, the Eigenvalues of the principal components are considered. The Eigenvalue of a principal component describes the variance accounted for by the component; the higher the Eigenvalue, the higher the variance in the data explained by the component.	
	The relationship is visually represented in the Scree Plot of Figure~\ref{fig:scree-plot} in the Appendix, which shows the Eigenvalue of each principal component for all components ordered by descending Eigenvalue. 
	To determine the optimal number of components to consider. \citet{scree-test} proposed to visually inspect the Scree Plot (\autoref{fig:scree-plot}) and determine the `elbow' in the graph where the plot levels off~\citep{scree-test}.
	In our case, this translates into considering the first three principal components; these $3$ components explained $43.22 \%$ of the total variance in the data. 
	
	We then considered the component loadings of the $19$ questionnaire items with the $3$ principal components, which quantify the correlation between a single item and the principal component.
	We removed $5$ items from further analysis due to their low component loading ($ <.5$), resulting in $14$ remaining items that accounted for $50.75$ of the total variance.% in the data. 

 	The resulting components and items appeared consistent (\autoref{tab:rotated-component-loadings}). The first component concerned performance and running goals such as pace or distance, confidence to reach one's goals, and feeling of freedom; we labeled it as \emph{performance \& achievement}. 
 	The second component reflected perceptions of the running environment in terms of beauty and cleanliness; we labeled it as \emph{perceived environment}. 
 	The last component captured a runner's feelings mainly in terms of freedom and social connectedness; we labeled it as \emph{mind \& social connectedness}. 
 	
\subsection{Experience of Running Scale (ERS)}
\label{sec:ers} 

\begin{table}[t!]
\footnotesize
% \small
\caption{ERS Questionnaire items for the pre- and post-run experience. The top part of the table (S1-S3) lists the short version of the questionnaire, which can be used standalone to capture the most relevant aspects according to the component analysis (\S\ref{subsec:components-running-experience}). For the long-form questionnaire, items L1--L13 can be used.} 
\label{tab:ers}
\begin{tabularx}{\textwidth}{ccXX}
\toprule
\# & Aspect & Pre-run Question & Post-run Question \\
\midrule
S1 & PA & How confident are you to reach your goal? & How easy was it to reach your goal? \\
S2 & ENV &  Are you happy with your environment right now?  & Was the route clean and beautiful? \\
S3 & MS & Do you feel connected to people?  & Do you feel connected to people?  \\
  \midrule
  \midrule
L1 & PA & How confident are you to reach your goal? & How easy was it to reach your goal? \\
L2 & PA & How confident are you to reach a certain pace? &  How easy was it to reach your intended pace? \\
L3 &PA & Do you think you will be able to feel free in your run?  & Did you have a feeling of freedom during your run? \\
L4 &PA & Do you anticipate frustration in your run? & Did you feel frustration during your run?\\
L5 &PA & Do you think you will be satisfied with this run? & All things considered, how satisfied were you with the run? \\
L6 &PA & Do you have specific expectations towards this run?& Was the running experience as you expected it to be?  \\ 
L7 & PA & How confident are you that you can run the distance? & How easy was it to run the distance? \\
L8 & ENV &  Are you happy with your environment right now? & Was the environment of your run nice? \\
L9 & ENV & Do you think the ground will be good to run on? &  Was the ground good to run on?   \\
L10 & ENV & Is the air quality good around you?  & Was the air quality of the route good?        \\
L11 & ENV & Are you pleased with the beauty of your surroundings? & Are you pleased with the beauty of the route?  \\ 
L12 & MS & Do you feel connected to people? & Do you feel connected to people?   \\
L13  & MS & Does your mind feel clear? & Does your mind feel clear?    \\
\midrule
\multicolumn{4}{l}{\textbf{PA:} performance \& achievement, \textbf{ENV:} environment, \textbf{MS:} mind \& social connectedness}\\
\bottomrule
\end{tabularx}
\end{table}	
 
Based on the extracted components of \emph{performance \& achievement}, \emph{ environment}, and \emph{mind \& social connectedness} and on the survey items belonging to these, we developed the questionnaire with three items for the pre- and post-run experience.
These six questions represent the minimum set of questions one should ask to capture the experience of a run~\citep{constantinides2020comfeel}. The questions with the highest component loadings (\autoref{tab:rotated-component-loadings}) per principal component were chosen to make our set of six questions (three questions before and after the run, respectively) for the ERS-short questionnaire (\autoref{tab:ers}).
	
The first component on \emph{performance \& achievement} contains questions regarding the goal (e.g., pace and distance) of the run and the confidence to achieve it. The item \emph{`It was easy to reach my running goal'} had the highest component loading and was chosen to represent this component in the post-run questionnaire, while the item \emph{`I am confident to reach my goal in this run'} is asked in the form of a question in the pre-run questionnaire. The second component, \emph{perceived environment}, characterizes questions on the environment, e.g., cleanliness or good running ground. The two items chosen to represent this component are \emph{`Are you happy with your environment right now?'} (pre-run) and \emph{`Was the route clean and beautiful?'} (post-run). The last component \emph{mind \& social connectedness} was represented by the item with the highest component loading for the pre- and post-run question: `I feel connected to people.'	

\subsection{Path Preferences}
\label{subsec:path_preferences}

\begin{figure}[t!]
\footnotesize
  \includegraphics[width=\textwidth]{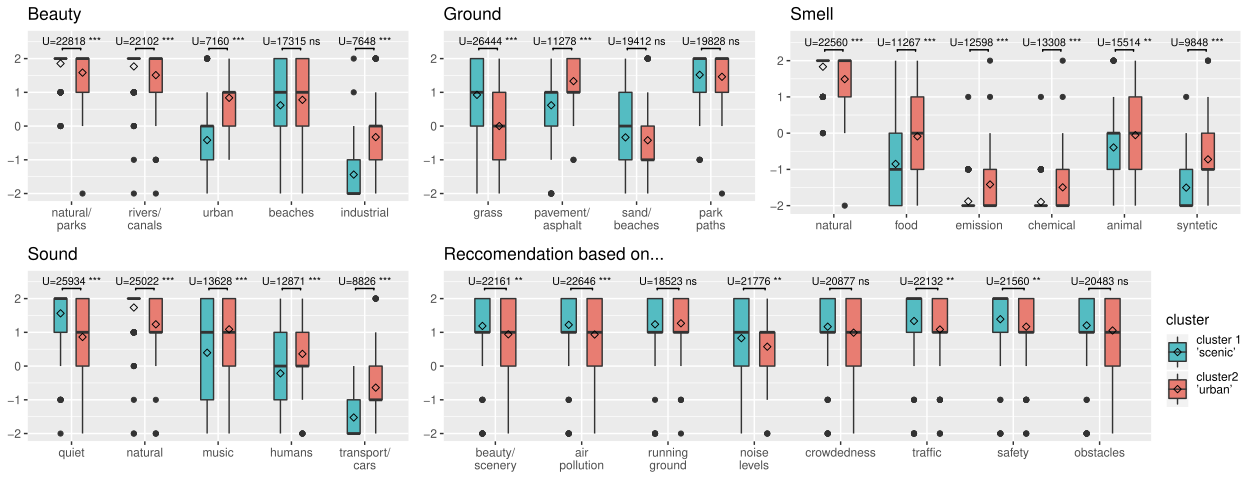}
  \caption[]{Environmental preferences of \emph{scenic} and \emph{urban} running paths. The plots show the preference ratings for each item (in the range of `totally dislike' (-2) to `totally like' (2)). Results of the Holm-Bonferroni corrected Mann-Whitney U comparison test are shown above each criterion (uncorrected, original $\alpha$ was `*' - $p < 0.05$, `**' - $p<0.01$, `***' - $p<0.001$.)}
  \label{fig:path-clusters}
\end{figure}

	After having studied the general running experience, let us focus on a specific aspect: path preferences. The survey contained 30 items related to this aspect, which captured the themes emerging from our previous interviews. We performed a $k$-means clustering over these  $30$ items. Based on Silhouette Analysis, we found that the best value for $k$ is $2$. The resulting clusters had a size of $157$ and $230$, respectively. Based on these clusters, we performed Mann-Whitney U tests over the $28$ features with a Holm-Bonferroni correction; a visualization of the clustered data and results of the pairwise Mann-Whitney U test can be found in \autoref{fig:path-clusters}. 

	The results of \autoref{fig:path-clusters} show that the two clusters were indeed orthogonal. We will refer to Cluster I as \emph{`scenic path'} as it represents respondents who prefer natural settings (they like greenery rather than routes full of people). 
	Further, we will refer to Cluster II as \emph{`urban path'} as it represents respondents who like urban settings (e.g., presence of people).

	Given previous studies that investigated personality and running habits~\citep{Sato2018Running,Chojnicki2021Key}, we hypothesized that personality might be a relevant factor. We assessed \emph{Openness}, \emph{Conscientiousness}, \emph{Extraversion}, \emph{Agreeableness}, and \emph{Neuroticism} of respondents with the BFI-10 Personality questionnaire~\citep{Rammstedt_BFI10}. 
	We performed the Mann-Whitney U tests to find differences between personality traits among our respondents; this is a non-parametric\footnote{The non-normality of the data was confirmed using Sapiro-Wilk test and visual inspection of the QQ-Plots.} alternative to the unpaired/independent t-test for testing the difference between two samples. 
	We found that the trait \emph{Neuroticism} was significantly ($U = 21100.5, p =0.0034$) higher for respondents who opted for \emph{scenic} paths ($\mu = 3.9 \pm2$) as opposed to the cluster with participants preferring \emph{urban} paths ($\mu = 3.4 \pm1.9$). 
	The opposite was true for the trait of \emph{Extraversion} ($U = 15649.5, p = 0.029$): extrovert participants opted for \emph{urban} paths ($\mu  = 4.2 \pm2.1$) rather than \emph{scenic} paths ($\mu  = 3.9 \pm2.1$).
	These differences highlight that there may be differences in path preference depending on personality traits.

\subsection{Additional Findings}

	At the end of the survey, we asked participants for additional comments on their use of technology and desirable features for a potential running app. We obtained $91$ replies, which we coded into four thematic groups. The first group concerned apps' tracking features. Elevation gain and terrain conditions were mentioned as desirable metrics to be tracked by $6$ respondents. Additionally, statistics such as pace, distance, and historical distance were mentioned as essential features. The second thematic group had to do with the integration of running apps with music players. Music is commonly used as a motivator for running~\citep{Bauer2015Designing}  (mentioned by $5$ respondents). In fact, Kim et al.~\citep{kim2020pepmusic} has demonstrated a computational way of recommending songs that are motivational for people given their contexts and activities by operationalizing the Brunel Music Rating Inventory (BMRI). The third group mentioned the importance of technologies that increase social connections in performance.
These technologies include the ability to virtual race against friends and awareness of where friends run at any given time.
	This is related to work from  \citet{Mueller-joggingdistance} to support a ``Jogging Together'' experience~\citep{Mueller-joggingdistance}. The fourth and final thematic group was about safety in its broader sense. Safety from crime was mentioned by 3 female respondents: \pquote{S43-f}{One of the things that stops me running is worrying I will be alone in a park.} Corresponding technological solutions included the display of emergency contacts in the lock screen (S30-m) and automatic notification of loved ones in case of a problem (S234-m). Safety from physical harm due to terrain conditions was also mentioned, and tripping hazards were reported as a desirable feature too. S29-m suggested community-sourced information on potential hazards and warnings. Finally, two participants proposed to display facilities that increase comfort, such as water fountains or toilets~(S15-f,S29-m).

\subsection{Summary}

The survey aimed to explore runners' sensory and environmental preferences, identifying three key components of the running experience: \emph{performance \& achievement}, \emph{perceived environment}, and \emph{mind \& social connectedness}.
These components formed the basis for the Experience of Running Scale (ERS), used to assess runners' experiences before and after their run.
Additionally, cluster analysis revealed two preference profiles among runners: \emph{scenic} and \emph{urban}. These profiles were correlated with personality traits, with Neuroticism linked to a preference for scenic experiences and Extraversion associated with a preference for urban environments.

\section{A Routine Engine for Scenic and Urban Paths} \label{sec:design}
Next, we provide an implementation of a routing engine that incorporates the path preferences previously identified, and then discuss the open challenges of designing such an engine. Recall, that the outcome of the cluster analysis (\S\ref{subsec:path_preferences}) indicated the presence of two main types of runners: those who prefer \emph{`scenic'} paths and those who prefer \emph{`urban'} ones.

\subsection{Street Weighting Scheme}

A geographical routing system takes in input a network of weighted street segments; then, given starting and destination points on any of those segments, the system selects an ordered sequence of connected segments that minimizes a given cumulative weight. Traditional routing systems aim at minimizing the traveled distance or travel time. To do so, to each segment, they assign weights based on the length of the segment or the time to travel it. 
To recommend custom routes, we designed a routing system that adapts those weights to consider the environmental factors in Figure~\ref{fig:path-clusters}, and the weights change depending on whether the path is \emph{scenic} or \emph{urban}. 
The information used in the weighting scheme is summarized in \autoref{tab:weighting}, and it is structured on a two-layer hierarchy. 
We computed street segment weight by linearly combining the street's values along the eight most important dimensions derived from the survey (e.g., beauty, absence of obstacles). To establish the relative importance of those dimensions, we then used their average rating from the survey. We use $\alpha_S^i$ and $\alpha_U^i$ to denote the average ratings for dimension $i$ for the \emph{scenic} ($S$) and \emph{urban} ($U$) paths, respectively. For example, the average rating for the importance of smell from respondents in the \emph{scenic} cluster was $\alpha^{smell}_S=1.50$. More generally, each segment $j$ was weighted as:

\begin{table*}[t!]
\centering
\footnotesize
\caption{The variables used in the routing weighting scheme. Seven dimensions are considered and combined linearly using the $\alpha$ coefficients. Some of the dimensions consist of sub-components, which are, in turn, combined linearly using the $\beta$ coefficients. All coefficients are extracted from the results of our survey and differ depending on whether the path to be recommended is \emph{scenic} ($\alpha_S$, $\beta_S$) or \emph{urban} ($\alpha_U$, $\beta_U$). The Proxy column lists the names of the variables (detailed in \S\ref{sec:design}).}

\begin{tabular}{crccl}
\toprule
 \textbf{Dimension} & \multicolumn{1}{c}{\textbf{Component}} & $\boldsymbol{\beta_S}$ & $\boldsymbol{\beta_U}$ &  \multicolumn{1}{c}{\textbf{Proxy}} \\ 
\midrule
 \multirow{7}{*}{\begin{tabular}{@{}c@{}}\emph{Smell} \\ $\alpha_S=1.50$ \\ $\alpha_U=0.40$\end{tabular}} & Nature    &  1.80 &  1.40 & $f^{smell}_{nature}$ \\
                         & Food      & -0.64 & -0.17 & $f^{smell}_{food}$ \\
                        & Emissions & -1.80 & -1.40 & $f^{emissions}_{food}$   \\
                         & Chemical  & -1.80 & -1.40 & $f^{chemical}_{food}$   \\
                         & Synthetic & -1.30 & -0.81 & $f^{smell}_{synth}$ \\
                         & Animals   & -0.23 & -0.19 & $f^{smell}_{animals}$\\
                         & Odorless  &  1.40 &  0.89 & --- \\
\midrule
 \multirow{5}{*}{\begin{tabular}{@{}c@{}}\emph{Sound} \\ $\alpha_S=1.20$ \\ $\alpha_U=-0.06$\end{tabular}}  & Natural   &  1.70 &  1.10 & $f^{sound}_{nature}$\\
                        & People    &  0.08 &  0.10 & $f^{sound}_{people}$\\
                        & Transport & -1.30 & -0.68 & $f^{sound}_{transport}$\\
                        & Music     &  0.81 &  0.67 & $f^{sound}_{music}$\\
                        & Quiet     &  1.40 &  0.89 & $f^{sound}_{quiet}$\\
\midrule
 \multirow{5}{*}{\begin{tabular}{@{}c@{}}\emph{Scenery} \\ $\alpha_S=1.50$ \\ $\alpha_U=0.42$\end{tabular}} & Natural & 1.90 & 1.50 &  \\
                         & River      &  1.80 &  1.40 & \\
                         & Urban      &  0.04 &  0.55 & beauty\\
                         & Beach      &  0.76 &  0.61 & \\
                         & Industrial & -1.10 & -0.43 & \\
\midrule
 \multirow{4}{*}{\begin{tabular}{@{}c@{}}\emph{Ground} \\ $\alpha_S=1.60$ \\ $\alpha_U=0.74$\end{tabular}} & Grass     &  0.58 &  0.22 & $OSM^{ground}_{grass}$\\
                         & Pavement  &  0.92 &  1.10 & $OSM^{ground}_{pavement}$\\
                         & Sand      & -0.37 & -0.39 & $OSM^{ground}_{sand}$\\
                         & Park      &  1.60 &  1.30 & $OSM^{ground}_{park}$\\
\midrule
 \multirow{3}{*}{\begin{tabular}{@{}c@{}}\emph{Obstacles} \\ $\alpha_S=1.60$ \\ $\alpha_U=0.39$\end{tabular}}       &       &    &    & \\
 & Absence of obstacles & --- & --- & $OSM_{obstacles}$\\
 & & & & \\
\midrule
 \multirow{3}{*}{\begin{tabular}{@{}c@{}}\emph{Traffic} \\ $\alpha_S=1.70$ \\ $\alpha_U=0.48$\end{tabular}}          &        &    &    & \\
 & Absence of traffic & --- & --- & $OSM_{way\_type}$\\
     &        &    &    & \\
%\midrule
% \multirow{3}{*}{\begin{tabular}{@{}c@{}}\emph{No crowdedness} \\ $\alpha_M=1.50$ \\ %$\alpha_V=0.37$\end{tabular}}          &        &    &    & \\
% & No crowdedness & --- & --- & ---\\
% & & & & \\
%\midrule
\midrule
 \multirow{3}{*}{\begin{tabular}{@{}c@{}}\emph{Safety} \\ $\alpha_S=1.70$ \\ $\alpha_U=0.58$\end{tabular}}                  &    &    & \\
 & Safety & --- & --- & Safety\\
 & & & & \\

\bottomrule
\end{tabular}

% \vspace{\tablecaptionspacing}

\label{tab:weighting}
\end{table*}

\begin{equation}
weight(j) = 1 / \sum^N_i(\alpha_{S \text{ or } U}^i \cdot dimension_i(j))
\label{eq:weighting_street}
\end{equation}

Where $N$ is the number of street segments, and $dimension_i(j)$ is the value of a dimension $i$ for the segment $j$.

Some dimensions consist of multiple sub-components. For example, the dimension capturing ground quality is broken down into four different types of pavement (i.e., grass, asphalt, sand, and park path). The relative importance of each dimension's component was also determined by the average rating of the corresponding survey items. We refer to those as $\beta^i_S$ and $\beta^i_U$ for the \emph{scenic} and \emph{urban} paths, respectively. A dimension score for a segment $j$ is therefore calculated as:
\begin{equation}
dimension_i(j) = \frac{1}{N} \cdot \sum^N_i(\beta_{S \text{ or } U}^i \cdot component_i(j))
\label{eq:weighting_segment}
\end{equation}
To make different dimensions (and components) comparable, we applied the log function to dimensions having skewed distributions and then mapped the resulting values in the range $[0,1]$ with a min-max normalization.  

\subsection{Street Weighting on Real Data}
\label{sec:street_weighting}

To apply the weighting scheme in practice, one needs to gather geo-referenced data that captures all our dimensions. We designed a proof-of-concept for Greater London.

\begin{description}[leftmargin=0cm]
	\item[Sensory perceptions.] 
	To gather smell and sound perceptions of each street in the city, we used the sensory mapping methodology proposed by~\citep{quercia2015-smellymaps,aiello2016-chattymaps}. They mined geo-referenced social media pictures to provide an estimation of the relative intensity of different types of smells and sounds that are likely to be perceived on the street. They then aggregated smells into 10 main categories and sounds into 8. These categories are the ones we considered in this work (listed in \autoref{tab:weighting}), with the exception of `odorless.' To then map the city's smellscapes and soundscapes, we used tags from 5.1M public Flickr picture tags taken in London\footnote{The data has been kindly provided upon request by \url{www.goodcitylife.org.}} over the course of 10 years, from 2005 to 2015. The value of each smell category $c$ was calculated as:
\begin{equation*}
f^{smell}_c(j) = \frac{\text{\#tags in segment $j$ of smell category $c$}}{\text{\#tags in segment $j$ of any smell category}}
\label{eq:f_smell}
\end{equation*}

After mapping sensory perceptions, we needed to compute paths. Using the same Flickr data, we used a  methodology similar to a previous work that estimated the aesthetic appeal of a street from geo-referenced pictures ~\citep{quercia2014-happymaps}. This approach accounted for the attractiveness of a street (if a spot was frequently photographed, it was likely to be visually interesting) and people's perceptions of a place (computed with the fraction of picture tags matching positive $f_p$ and negative $f_n$ emotion words in LIWC~\citep{liwc}, a standard dictionary of English words that reflect people's emotional and cognitive perceptions). More specifically, for a segment $j$, we calculated:
\begin{equation*}
beauty(j) = 0.03 \cdot log(\text{\#pics in $j$}) + 0.20 \cdot f_p(j) - 0.21 \cdot f_n(j)
\label{eq:beauty}
\end{equation*}

	\item[Street qualities.]
	Ground conditions, absence of obstacles,  and average traffic flow are all properties that are strongly linked to street type. 
	OpenStreetMap\footnote{\url{https://www.openstreetmap.org}} (OSM) is a convenient tool to gather street information for thousands of cities around the world.  
%Street segments in OSM are marked with key:value pairs that encode their different qualities.
We used the surface type of path segments to determine their suitability for running. These are binary indicators that reflect whether a street segment is of surface type $t$ --- referred to as $OSM^{ground}_{t}$ in \autoref{tab:weighting}.
If a street segment had a known surface, we counted it with a  $\beta$ weight that depended on the surface type.
To then model the absence of obstacles ($OSM_{obstacles}$ in \autoref{tab:weighting}), we computed the inverse count for traffic signals, stop signals, and give-way signals (all available as annotations of the street segments on OSM). 

OSM also encodes the type of road, ranging from \emph{motorways} to countryside \emph{tracks}, which can be used to estimate the average traffic flow on that road. We grouped those values in three sets reflecting increasing traffic flows: low traffic, high traffic, and extreme traffic. In addition to that, we used the OSM \emph{sidewalk} label to determine the number of sidewalks present on the road. A segment was assigned a score $OSM_{way\_type}$, which is highest for low-traffic roads and lowest for high-traffic streets. Furthermore, streets with extreme traffic and those with high traffic and no sidewalk were removed and never considered by the routing algorithm.

	\item[Safety.]
Official crime data is publicly available for the city of London. We queried the API of the crime and policing portal of England\footnote{\url{https://data.police.uk/docs/method/crime-street/}} to retrieve, for each street segment, the metadata on the crimes committed in 2018 within a 200m buffer around the segment. Our \emph{safety} score was then the inverse count of the number of crimes against a person (e.g., violent crimes, robbery) committed in each buffer area.

\end{description}

\subsection{Routing Engine}

To calculate trajectories, we loaded the OSM street network of Greater London and the modified street segment weights into GraphHopper,\footnote{\url{https://www.graphhopper.com/}} a routing engine suited for pedestrian routing. We extended GraphHopper to compute the \revision{best} path between two points using weights that changed depending on whether the path to be computed was \emph{scenic} or \emph{urban}. \revision{The routing engine trades-off the detour taken for achieving better weights with the detour needed. Algorithmically speaking, the cost of traveling on a path that is more in line with the requested running profile is decreased.} To simulate a typical running scenario, we generated circular paths (paths for which starting and destination points coincide) of user-defined length. To achieve that, GraphHopper uses the A* algorithm~\citep{goldberg2005computing}. It computes two non-overlapping paths between the starting point and an intermediate point and then back to the starting point. The intermediate point was set according to the total distance and a heading parameter (which characterizes path direction). In our implementation, we uniformly sampled $k$ headings, computed paths for all of them, and returned the one with the \revision{best score for the respective path type}. \revision{To ensure reproducibility, we have open-sourced the repository on GitHub\footnote{\url{https://github.com/rschifan/graphhopper}}, including the weighting for the OSM segment IDs.}

\begin{figure}[t!]
\centering
	\includegraphics[width=\linewidth]{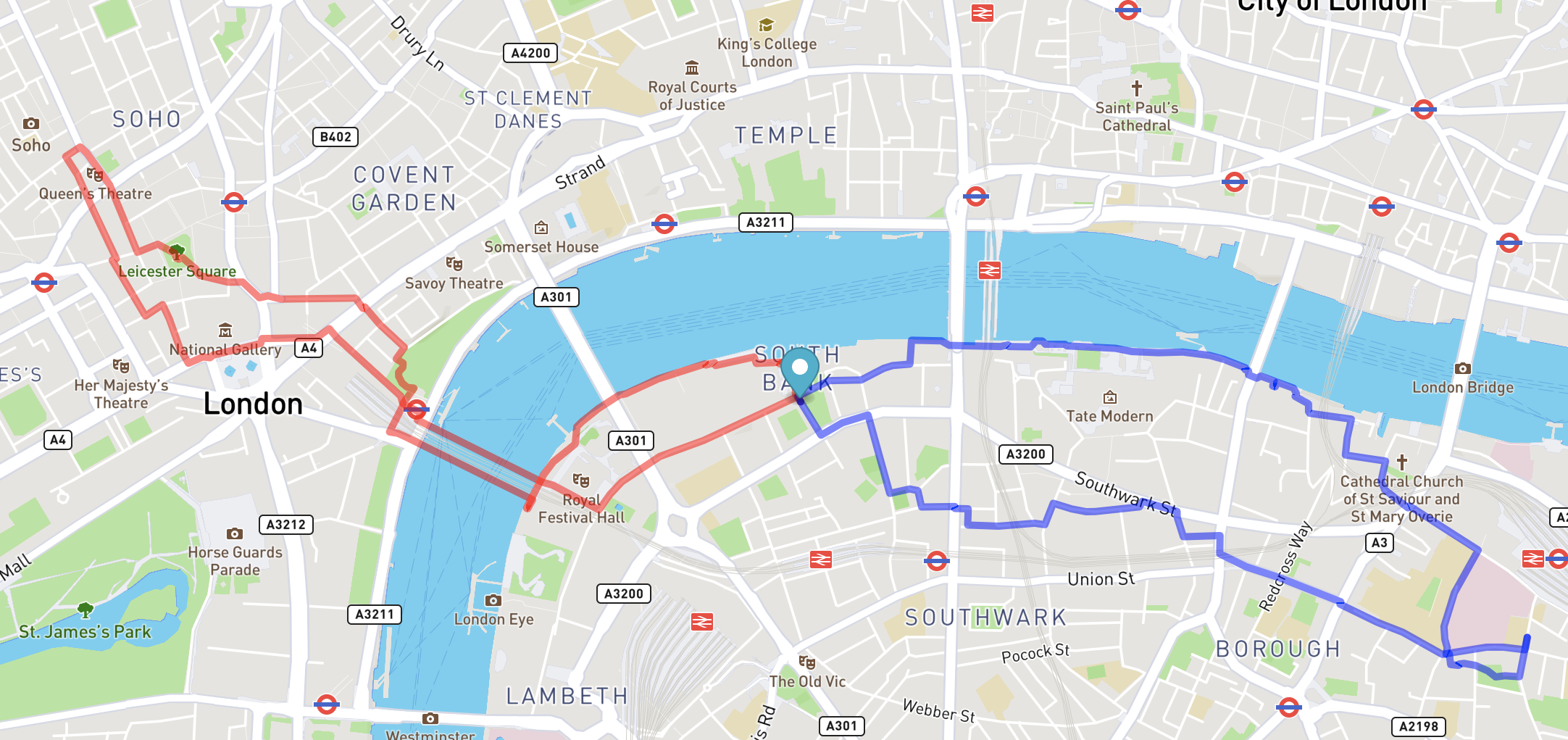}
	\caption{Example showing \emph{scenic} (blue, eastwards route) and \emph{urban} (red, westwards route) paths of 5km length in central London.}
	\label{fig:sample-path}
\end{figure}

A sample of the two routing schemes can be seen in \autoref{fig:sample-path}. The two circular paths were generated with a starting point in Southbank (central London) and had a length of 5 km. The blue (eastward) path represents a \emph{scenic} route, and the red (westwards) path is an \emph{urban} route. The \emph{scenic} path is more scenic and quieter as it is along the river. The \emph{urban} path steers toward Trafalgar Square and Soho (lively areas in central London).

\subsection{Quantitative Analysis of Routing Characteristics}

\begin{figure}[t]
    \centering
     \begin{subfigure}[t]{0.5\textwidth}
        \centering
        \includegraphics[width=\linewidth]{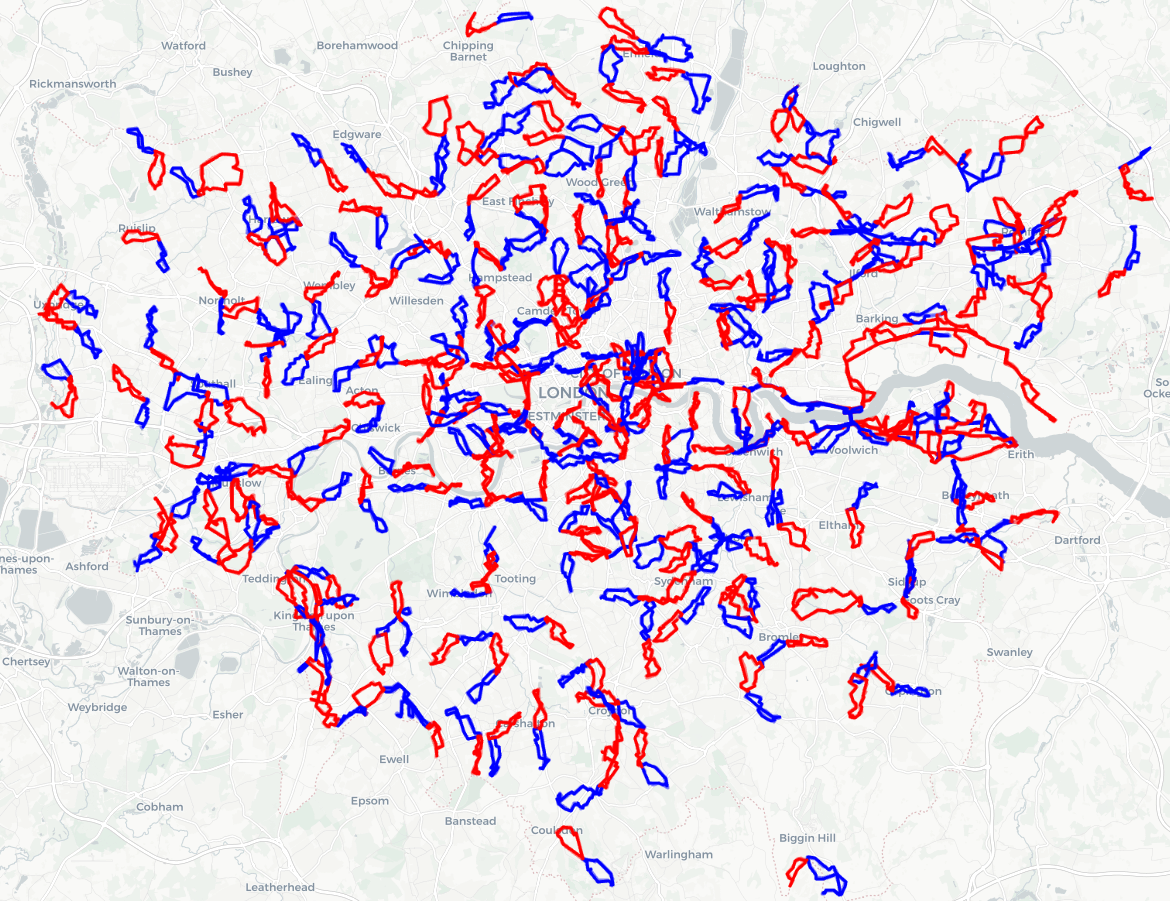}
        \caption{\revision{Route Coverage in London. The red paths correspond to urban, the blue ones to scenic routes.}}
        \label{fig:coverage}
    \end{subfigure}%
    ~ 
    \begin{subfigure}[t]{.5\textwidth}
        \centering
        \includegraphics[width=\linewidth]{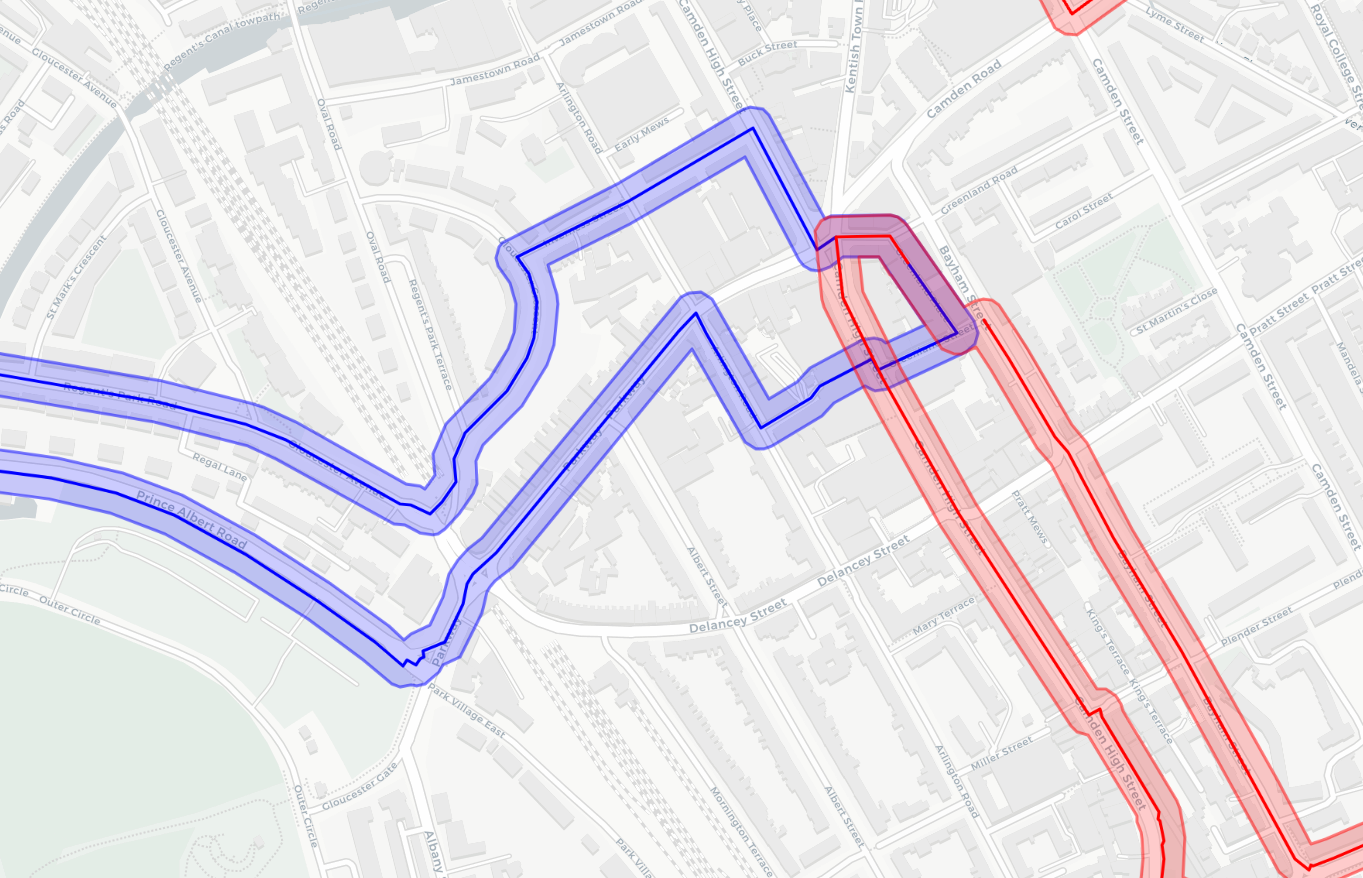}
        \caption{\revision{Defining a 25m buffer (shaded area) around the routes for querying images.}}
        \label{fig:buffer}
    \end{subfigure}
    \caption{\revision{(a) Coverage of evaluated routes, and (b) the area used to characterize those routes using Flickr images.}}
    \label{fig:validation_steps}
\end{figure}

\revision{To explore potential differences in the running experience between the two route types, we conducted a systematic analysis of 446 generated paths across London. For uniform coverage across the city, we used postal areas as query points in the routing engine, generating both a scenic and an urban round-trip route of approximately 5 kilometers each—equivalent to a 25-35 minute run for a recreational runner, as shown in \autoref{fig:coverage}.}

\revision{
To analyze the characteristics of different route types, we used the original Flickr dataset of georeferenced images in London (see \autoref{sec:street_weighting}.
These images are annotated with tags generated by a computer vision algorithm \citep{Li2009Towards,Thomee2016YFCC100M}.
To identify images associated with each route, we applied a 25-meter buffer around the paths, representing the immediate vicinity of the routes \autoref{fig:buffer}).
This 25-meter boundary was chosen after visually inspecting buffers of various distances, ranging from 10 to 50 meters. We found 25 meters to be the maximum distance typically observable while running in dense urban areas. Although some routes, such as riverside paths, offer more distant views, increasing the buffer size would capture backyards and other areas not visible to runners.}

\revision{
Querying the images within each route's buffer produced a collection of computer vision tags describing their content. For each route, we calculated the relative frequency of each label to normalize the varying numbers of images across route types. Finally, we computed the ratio of these frequencies between scenic and urban routes to determine the relative importance of each tag. This approach, formalized in \autoref{eq:importance},  helped mitigate biases due to differences in image density across city areas and route types.
}
\begin{equation}
importance_t = \frac{\sum_{s \in \{scenic~routes\}} 1 / N\frac{f_{t \in s}}{n_s}}{\sum_{u \in \{urban~routes\}} 1 / N\frac{f_{t \in u}}{n_u}},
\label{eq:importance}
\end{equation}
\revision{where $t$ is a tag, $N$ is the number of route pairs, $f_t$ is the frequency of the tag in a route and $n_s$ is the number of tags in the respective route.}

\begin{figure}
    \centering
    \includegraphics[width=\linewidth]{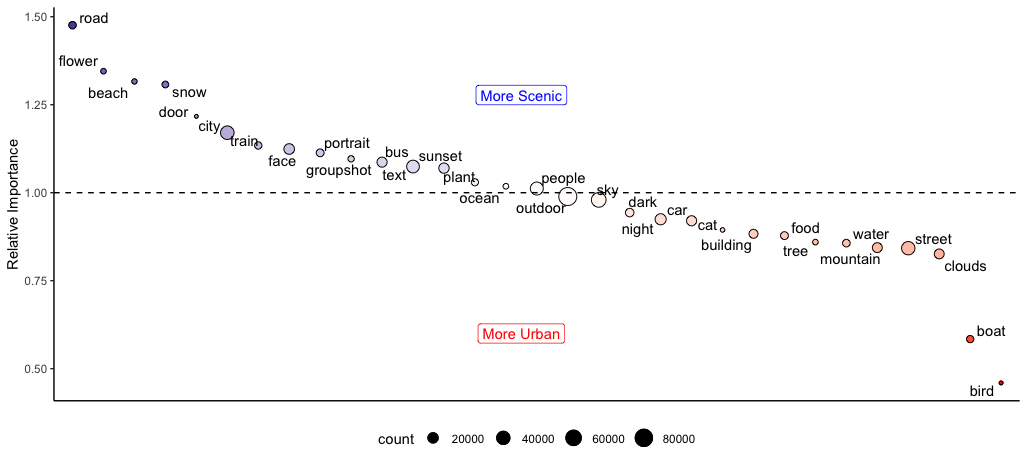}
    \caption{\revision{The relative importance of computer vision tags along the routes, with tags positioned higher on the vertical axis appearing more frequently in scenic routes compared to urban ones. Tags with a relative importance close to 1 are equally represented in both route types. Circle size corresponds to the number of occurrences of each tag.}}
    \label{fig:validation}
\end{figure}

\revision{\autoref{fig:validation} shows the outcome of the analysis for vision tags that occurred more than 1500 times to avoid effects stemming from small number of tags.
Unsurprisingly, generic and frequent tags, such as ``people,'' ``sky,'' and ``outdoor'' have a similar frequency in both route types.
In the top-left corner, we see tags that occur predominantly in \emph{scenic} routes, such as ``road,'' ``flower,'' ``beach,'' and ``snow.''
These tags are contrasted in the bottom-right corner of the most \emph{urban} tags: ``bird,'' ``boat,'' ``clouds,'' and ``street''.}

\revision{Reflecting on the relative frequencies of these tags, we made the following observations: 
(1) A considerable proportion of generic tags such as ``people'', ``outdoor'', and ``sky'' are equally distributed, which is expected given that the routes are all computed within a dense urban area and given that route pairs have the same starting point, there is some overlap between the routes.
(2) The highly \emph{scenic} tags depict concepts that are in line with the route type, such as flowers, beach (used in the tagging as a aggregate for any transition from water to land), ``snow'', and the ``city'' itself. We conclude that the tag ``city'' appears more frequently in scenic routes, as urban routes in more densely built-up areas do not allow for scenic overviews of the city, i.e., the skyline. This observation is further supported by the difference of ``road'' (\emph{a long, hard surface built for vehicles to travel along} \citep{CambridgeDict}--$1.48$ on the scenic spectrum) and ``street'' (\emph{``a road in a city or town that has buildings that are usually close together along one or both sides''} \citep{CambridgeDict}--$0.84$ on the urban spectrum).
(3) On the urban routes, we observed tags like ``building'', ``food'', ``street'', and ``bird.'' We attribute the relatively high frequency of ``bird'' tags to pigeons, which are more commonly photographed up close on bustling streets than in quieter areas.
}

\subsection{Qualitative Validation of Routing Characteristics}
\revision{To further validate the routing characteristics, we conducted interviews with nine runners based in London. Our sample included six men and three women, aged 25 to 48, with weekly mileage ranging from 5 to 70 km and running experience spanning 2 to 10 years. 
Participants were recruited through a pre-screening survey in London sports clubs, where they provided their home and work locations at the three-letter postcode level to ensure anonymity. Using these locations, we generated routes tailored to each participant's local context. Additionally, we included a third location near Canary Wharf to offer a common point of comparison, as Canary Wharf is one of the best-known areas outside London’s tourist center. Each participant received a set of routes, consisting of three urban and three scenic paths across the selected locations.
Following the presentations, we conducted the interviews to capture participants' preferences, reasoning, and perceptions of the routing characteristics. We report next the key themes emerged by thematically analyzing the interviews.}
\smallskip

\noindent\textbf{Preference for Scenic and Quiet Routes.} \revision{Participants consistently showed a preference for scenic routes, particularly those with natural elements such as canals, parks, and quieter streets. These routes were described as more enjoyable and visually pleasing, often offering an escape from urban noise and congestion. For example, one participant expressed a clear preference for scenic routes, saying, \emph{``I would prefer to run the blue one because it goes along the canal, and I really like the canal; it's super beautiful''} (P6). Similarly, another participant favored quieter, less trafficked paths: \emph{``I would take the blue... it's by the river, quieter from cars and public transport''} (P1). Some participants also highlighted that certain segments of scenic routes, particularly those intersecting busy roads, could detract from the overall experience. As one noted, \emph{``I don't like that part [along the main road], but I'd still take it over a busier, more urban route''} (P5).}
\smallskip

\noindent\textbf{Variety and Circularity in Route Design.} \revision{Many participants preferred routes with variety, particularly those that offered circular or non-repetitive layouts. Routes that avoided backtracking were perceived as more engaging, with the opportunity to experience new sights along the way. One participant described their preference for circular routes, stating that, \emph{``the scenic routes don't make you feel like you're running the same path twice, which is nice for staying motivated''} (P4). Similarly, P7 disliked routes with excessive turns, describing them as ``wiggly'' and interruptive to their flow. These comments underline the importance of route design in maintaining participants’ interest and motivation, especially during longer runs.}
\smallskip

\noindent\textbf{Concerns with Urban Routes.} \revision{Urban routes were generally less favored by participants, who often found them too busy, visually unappealing, or even challenging in terms of air quality and pedestrian congestion. For example, one participant expressed a strong dislike for Finchley Road, saying, \emph{``I hate Finchley Road; it's awful for running''} (P2). Another participant shared that urban routes often follow busier streets with ``city traffic'', which felt less ideal for running (P1). Some participants were also disappointed that the routing engine did not prioritize more visually appealing urban landscapes, with P9 suggesting, \emph{``if it's an urban run, it would be more interesting around landmarks or quieter streets''}. This feedback highlights the challenge of designing urban routes that are both accessible and enjoyable for runners.}
\smallskip

\noindent\textbf{Safety Considerations and Night Running.} \revision{Safety was a significant consideration for many participants, particularly when planning runs in the evening or at night. Well-lit, busier urban routes were often perceived as safer options for night running, whereas scenic routes near isolated areas such as canals were viewed as potentially risky. P6 highlighted this concern, saying that, \emph{``if I were running at night, I'd definitely go with the urban one...[the scenic] feels isolated, and I wouldn't feel comfortable''}. P3 also expressed concerns about certain paths, stating that they would avoid scenic routes at night due to the potential lack of visibility and foot traffic. This situational aspect of route choice highlights how safety concerns can impact route preference, especially based on the time of day.}
\smallskip

\noindent\textbf{Navigation Challenges and Route Layout.} \revision{Participants expressed frustration with routes that featured unnecessary turns or complex layouts, which required more focus and disrupted their running rhythm. For example, P2 criticized the layout of some routes, explaining, \emph{``the scenic has some funny turns... I'd prefer a more straightforward path that lets me focus on my run''}. P5 echoed this frustration, stating that routes with frequent turns required more effort to navigate (\emph{``the urban route feels more challenging to navigate with all the turns''}). Routes with simpler and less directionally complex layouts were preferred, as they allowed runners to achieve a ``flow'' state without being disrupted by navigation issues.}
\smallskip

\noindent\textbf{Familiarity with Area and Route Suitability.} \revision{The degree of familiarity with a given area influenced participants' route choices, with known areas offering comfort and predictability. P9 described their familiarity with running near Brockwell Park, stating that they feel more comfortable running in well-known areas, especially at night: \emph{``I do a very similar run to the Herne Hill one at night... it's one of my dark runs''} (P9). P1 similarly preferred familiar routes, explaining, \emph{``I know this part of town, so I feel better choosing it---even if it's urban, I'm familiar with where to avoid the worst parts''}. Participants also indicated that their preferences could change depending on the season or time of day, with one runner explaining, \emph{``in winter, when it gets dark early, I stick to areas I know and places with more people around''}~(P4).}

\subsection{Open Challenges in Designing the Routing Engine}

	Our approach to determining routing weights relied on data gathered from an online survey, particularly focusing on respondents' perceptions about environmental factors affecting their running experience. While this methodology represents a notable advancement over conventional routing engines, which often consider only singular aspects like ground type, it remains a static representation of the running environment.
	To enhance its utility further, we propose making the routing adaptive to various environmental contexts, including weather conditions, seasons, and the specific goals or intentions behind each run. By incorporating these dynamic variables, we can significantly broaden the scope of route recommendations and improve their personalization.
	However, the main challenge of performing a large-scale user study to prove effectiveness remains for future work.

\section{Discussion}
\label{sec:discussion}

The primary conclusion of this work is that the running experience is highly personal. 
Our interviews and survey showed that judgments on what constitutes the optimal route differ across individuals.
Emerging contrasts, e.g., goal-driven \emph{vs.} recreational, natural \emph{vs.} fun-urban experiences, or exploratory \emph{vs.} familiar routes show the multi-faceted nature of running.
The orthogonal components uncovered in the quantitative analysis, namely \emph{performance \& achievements}, \emph{environment}, and \emph{mind \& social connectedness}, reflect this nature. 

\subsection{Main Findings} 
\label{sec:main_findings}

Based on a mixed-method approach of semi-structured interviews and a subsequent online survey, we shed light on which aspects are relevant for the \textit{Experience of Running}. 
In the qualitative interview study, we were able to derive 5 themes of the running experience, which we summarized as ``Abilities,'' ``Path choice strategies,'' ``Environmental,'' ``Focus,'' and ``Post-Run.''
These themes provide a comprehensive picture of what should be considered when designing technologies for runners; from planning the route before the run to perceiving the environment and dealing with the physical exertion during the run, and to the post-run experience after the run. Although these themes were derived from 7 participants, they proved to be useful in informing the design of the online survey for a more generalized exploration of the topic.

The online survey with 387 participants, in turn, revealed three components: \emph{performance \& goals}, \emph{environment}, and \emph{mind \& social connectedness}, which informed the design of the Experience of Running Scale~(ERS). Using the responses in the quantitative survey, we clustered the path preferences of users and uncovered that there are two major groups: \textit{``urban''} and \textit{``scenic.''}
After verifying that there are differences in path preferences depending on personality traits, we used sensory mapping of urban running paths to provide computationally-generated running routes that fit personal preferences and current needs.
Instead of offering a fully customizable experience, we focus on these two orthogonal choices, which are meaningful (as they emerged from our quantitative analysis) and promise to avoid issues revolving around choice overload~\citep{haynes-choice-overload}.
Furthermore, these preference profiles are directly applicable for the routing algorithm, which is ususally not the case in previous characterizations of runners~\citep{Besomi2018SeRUN,Ogles2003typology}.

As opposed to the common assumption runners all value green spaces~\citep{Deelen2019Attractive}, we found that there is a large proportion of runners who value urban settings and use running as a tool to explore new areas around their homes. 
One emerging facet of our interviews and survey was that the presence of people was a contrasting source of either frustration due to overcrowding but also a feeling of safety when running in deserted areas.
\revision{These considerations were supported by the qualitative evaluation of the generated routes, which indicated a general preference among runners for scenic routes. However, runners tended to rely more on urban routes in the evening or when navigating unfamiliar areas.}

\subsection{Implications}
From a theoretical standpoint, we set the foundation for capturing the previously under-investigated concept of the ``Experience of Running'' for recreational runners.
We revealed that the focus of many fitness tracking apps on the ``performance \& achievement'' aspect, i.e., the athletic goal of a run, is only a small portion of the larger picture, which should be addressed in the future.
We developed the ERS Questionnaire as a very compact 3-item questionnaire that captures the most relevant aspects of the running experience (cf. \S\ref{sec:ers}).
It provides a quick yet comprehensive pre- and post-assessment of the runners' state of mind regarding the run.
The brevity is by choice to cater to the reality of specialized devices, where the interaction is dictated by constraints (e.g., the display of a sports watch).
The ERS responses can provide a more nuanced view of the running experience compared to a self-assessment regarding the perceived effort or the `feel' as currently done after finishing a workout on various devices such as those from Garmin~\citep{Repici2019Because}.
The questionnaire can also prove useful in the assessment of derived metrics about the run~\citep{Bentvelzen2023Designing}.

From a practical perspective, we showcased that it is feasible to use automated ways to generate a rich image of the urban environment for runners, and transform this into weights for street segments to be used as input for running-specific routing engines.
Thus, we were able to extend the toolbox for route generation with ``urban'' and ``scenic'' routing options, which previously were limited to basic aspects, such as the underground, hilliness, or popularity.

Our work sets the foundation for incorporating novel route options that go beyond the shortest or the popular paths, or simple features such as the running surface.
While our weighting scheme in the route generation was dictated by the answers of a relatively large sample recruited from Facebook, the weights can be elicitepd in an alternative way.
This opens the possibility to even personalize the weights using user modeling and preference elicitation methods~\citep{kassak2015user}.

\subsection{Limitations and Future Work}
Our work comes with \revision{five} limitations that call for future research. \revision{The first limitation is about the relatively small qualitative sample size (7 participants). While we reached thematic saturation during the interviews, and the small sample was sufficient for exploratory purposes, a larger sample could have provided a broader range of experiences. However, we mitigated this limitation by complementing the qualitative findings with a large-scale quantitative survey (387 respondents), which allowed for a more generalizable understanding of running experiences.}

\revision{The second limitation relates to the demographic homogeneity of the survey participants. The majority of participants were aged 45-54 years (32.8\%) and identified as either male or female, with minimal representation from non-binary or other gender identities. This may limit the generalizability of our findings, as runners from different age groups, particularly younger or older participants, as well as those from more diverse gender backgrounds, may have different experiences and preferences that are not fully captured in this study. Future research should aim to recruit a more diverse participant pool to explore whether demographic factors significantly influence the running experience.}

\revision{The third limitation is the reliance on self-reported data, which introduces potential biases such as recall bias and social desirability bias. Recall bias may have affected participants' accuracy in reporting past runs, particularly if significant time elapsed between the running experience and the survey response. Although we focused on participants' most recent run, memory distortions may still influence the precision of their responses. Additionally, social desirability bias could have led some participants to provide responses they perceived as more favorable or acceptable, especially in terms of running performance or preferences. While self-reported data is valuable for capturing subjective experiences, future studies could complement these findings with objective data (e.g., GPS tracking, biometric sensors) to mitigate these biases. Additionally, the weights assigned to different routes may impact the quality of the generated routes. However, conducting an ablation study by adjusting route weights requires ground truth information on the ``optimal routes'' and can be computationally costly.}

The \revision{fourth} limitation is about the availability of fine-grained data to capture specific sensory aspects of the city contributing to the running experience. Here, we see air quality information as the most relevant missing piece, which could not be addressed because of the unavailability of granular air pollution data. We still point out that we partly addressed this with the smellscape features (Table~\ref{tab:weighting}), while recently proposed approaches could be employed to improve this aspect~\citep{Chen2019Deep,Xia2019Revealing}. 
Additionally, incorporating human mobility analysis into our model could enhance our understanding of spatio-temporal variations in route safety~\citep{Wang2022Spatio}. \revision{Moreover, to enhance the responsiveness and context-awareness of the routing engine, integrating real-time data such as weather conditions, time of day, and crowd density~\citep{hillen2015routing} would be beneficial. For example, routes could be dynamically adjusted to avoid congested areas or optimized for safety during nighttime runs.} It is also important to acknowledge the difficulty in capturing temporally changing aspects of the environment. Most importantly, social aspects of running in a group profoundly influence the running experience, but cannot be estimated using available data sources.
The HCI community addressed  social aspects in the context of running by augmenting and creating a shared experience between distant runners~\citep{Mueller-joggingdistance,Alohali2016Run}, offering artificial companions (e.g., drones)~\citep{Mueller-drone}, or facilitating support between runners and spectators during races~\citep{Wozniak-rufus}. 

The \revision{fifth} limitation concerns the focus of the sensory mapping of only one city, that is, London, United Kingdom. The choice of the British capital was due to data availability of prior work~\citep{aiello2016-chattymaps,quercia2015-smellymaps}, and its substantial size as a metropolitan city, serving the purposes of our feasibility study. Future studies could replicate our methodology in other European or Western metropolitan cities upon data availability.

Future work should focus on empirically quantifying the interplay between route choice, perceived running experience, contextual information (e.g., time of day, weather), and sensed data (e.g., running speed and heart rate). We believe that such data can be leveraged to develop improved recommendation models for running routes and provide insights into user needs in different contexts~\citep{Baldauf2007Survey,Adomavicius2015Context}.
Another promising direction is to explore running experience through the lens of psychological theory. Association and dissociation~\citep{Morgan-associ}, two concepts from behavioral psychology, emerged as coping strategies that were mentioned by our interview participants (e.g., listening to music or attending to bodily sensations).
Although our study could not directly address this, future running apps could enhance user experience by understanding a runner's mental state through pre-run questions (e.g., captured by our ERS scale) and sensor data.

\section{Conclusion}

This study highlights the environmental and sensory factors shaping the overall experience of runners in urban environments.
Existing methods for generating running routes, based on popularity or the running surface, have limitations in urban areas and lack personalization.
To address these limitations, we proposed an approach using computational running routes based on sensory mapping to approximate users' subjective experiences and demonstrated their applicability in a routing engine for running paths recommendations.
Our contributions include identifying key themes from interviews, developing the Experience of Running Scale (ERS) to measure the pre- and post-run experiences, and extracting and instantiating two path preferences associated with different experiences, namely ``urban'' and ``scenic'' route options. This research provides a starting point for further research in the development of more customized and enjoyable running routes that promote physical activity and safety in urban environments.



\bibliographystyle{elsarticle-harv} 
\bibliography{main}

\appendix
\clearpage

\section*{Appendix}

\begin{table}[htb]
	\caption{Interview topics and question guidelines for the semi-structured interviews.}
	\label{tab:interview-guide}
        \scalebox{0.7}{
	\begin{tabular}{ll}
		\toprule
		\multicolumn{1}{l}{\textbf{Topic}}                                                   & \multicolumn{1}{l}{\textbf{Questions Guide} \& \textbf{Prompts}}                                                                                                                                                                                                                                                                                                                                          \\ \midrule
		\textbf{Motivation}                                                                           & \begin{tabular}[t]{@{}l@{}}- Can you tell me a bit about the motivation for running?\\ - How often do you go running?\end{tabular}                                                                                                                                                                                                                                               \vspace{.5em}\\
		\textbf{Preparations}                                                                         & \begin{tabular}[t]{@{}l@{}}- What preparations do you make before you go for a run?\\ - How do you find a good route for your run?\\ - What are the criteria for the optimal route?\end{tabular}                                                                                                                                                                           \vspace{.5em}\\
		\textbf{Experience}                                                           & \begin{tabular}[t]{@{}l@{}}- What are the typical things you do and experience before the run?\\ - On a run, what are the things you usually experience during a run?\\ - Which environmental aspects are important when you are on a run? \\- How important is the quality of the urban space for you during the run?\\ - What are the typical things you do and experience after the run?\end{tabular} 
		\vspace{.5em}\\
		\begin{tabular}[t]{@{}r@{}}\textbf{Technology \& Reflection}\end{tabular} & \begin{tabular}[t]{@{}l@{}}- Do you use any technology when you go for a run? How/why do you use it?\\ - What are the other important aspects of your run that current technology \\ \hspace{1em} does not capture? \\ - What support is useful in terms of reflection or future technologies?\end{tabular}                                                                                                        \\ \bottomrule
	\end{tabular}
 }
\end{table}

\vspace{30pt}

\begin{figure*}[htb]
	\includegraphics[width=0.9\textwidth]{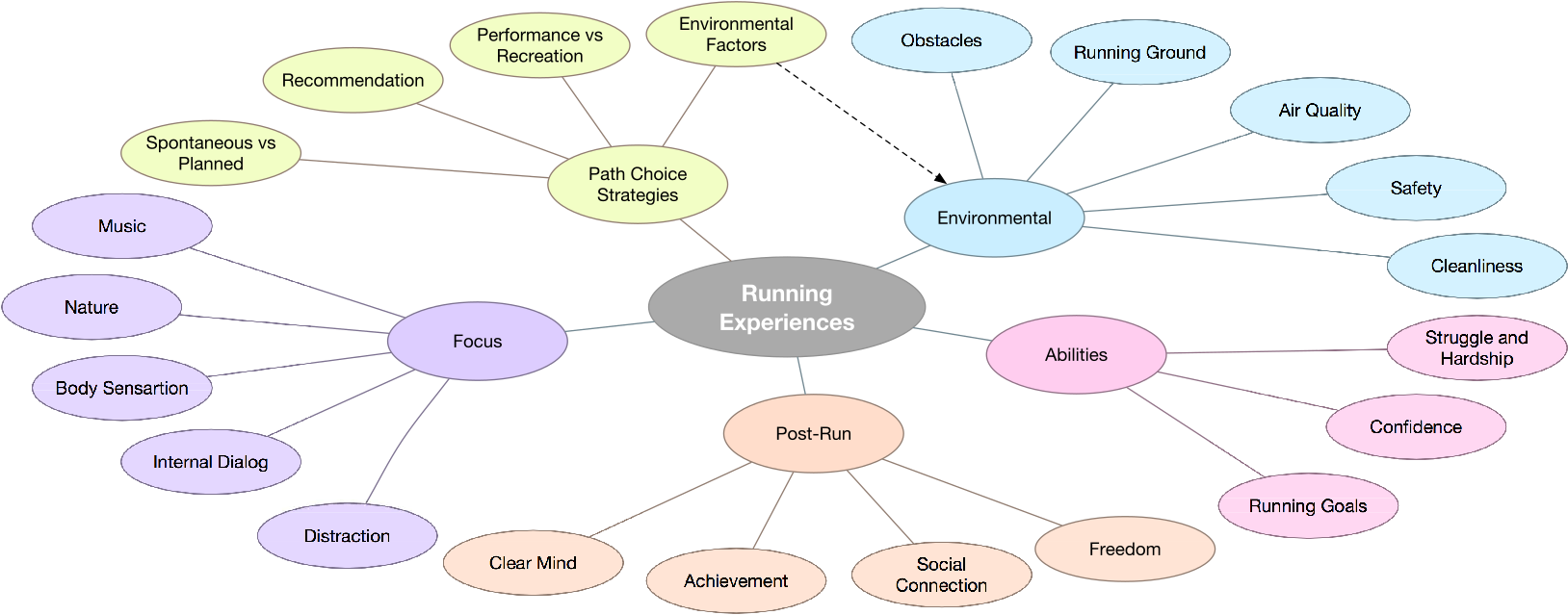}
	\caption{Overview of the main themes and sub-themes from the interviews.}
	\label{fig:experience-themes}
\end{figure*}

\begin{table}[htb]
	\sffamily
	% \small
 \caption{Demographic information of survey participants.}
	\label{tab:survey-demographics}
	\begin{subtable}[c]{.4\linewidth}
		\centering
		\begin{tabular}{@{}rcc@{}}
			\toprule
			& \#answers & \%\\ \midrule
			18-24 years old   & 28      & 7.2\\
			25-34 years old   & 67      & 17.3\\
			35-44 years old   & 79      & 20.4\\
			45-54 years old   & 127     & 32.8\\
			55-64 years old   & 62      & 16.0\\
			65 years or older & 24      & 6.3\\ \bottomrule
		\end{tabular}
		\vspace{\tablecaptionspacing}
		\caption{Age distribution}
		\label{tab:survey-demographics-age}
	\end{subtable}
	\quad
	\quad
	\begin{subtable}[c]{.4\linewidth}
		\centering
		\begin{tabular}{@{}rcc@{}}
			\toprule
			& \#answers & \% \\ \midrule
			female                   & 203     & 52.5\\
			male                     & 180     & 46.5\\
			non-binary               & 1       & 0.2\\
			other                    & 1       & 0.2\\
			not disclosed & 2      & 0.5 \\ 
			& \\
			\bottomrule
		\end{tabular}
		\vspace{\tablecaptionspacing}
		\caption{Gender distribution}
		\label{tab:survey-demographics-gender}
	\end{subtable}

\end{table}

	\begin{figure*}[t!]
	\includegraphics[width = 0.9\linewidth]{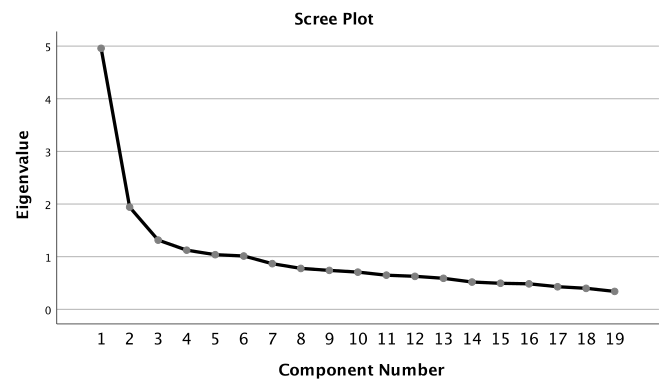}
	\caption{The Scree Plot visualizes the Eigenvalues related to each principal component. The components are ordered by descending Eigenvalue. This means Component 1, with the highest Eigenvalue, captures the most variation in the data, Component 2 the second most, and so on.}
	\label{fig:scree-plot}	
\end{figure*}




% \end{thebibliography}
\end{document}